\documentclass[12pt,letterpaper]{article}
\usepackage{epsfig,pslatex}

\begin{document}

\begin{titlepage}

\vspace*{1cm}

\begin{center}
{\bf \Large  Angular Correlations in Associated Production of Single
    Top and Higgs with and without anomalous $Wtb$ Couplings}
\vspace{2.cm}
\end{center}

\begin{center}

{\bf S. F. Taghavi and M. Mohammadi Najafabadi} \\
\vspace{0.1cm}
{School of Particles and Accelerators, Institute for Research in
  Fundamental Sciences (IPM), P.O. Box 19395-5531, Tehran, Iran} \\

\end{center}

\vspace{1.5cm}

\begin{center}
{\bf Abstract}
\end{center}
We study the angular correlation and the amount of top quark
polarization in the production of a higgs boson in association with
a single top quark in the $t-$channel at the LHC.
We also study the effect of anomalous $Wtb$ couplings on the
angular correlation and on the production cross section of the process.
The cross section and angular correlation is almost insensitive to the
variation of the Higgs boson mass within 3 GeV.
The robustness of the angular correlation against the center-of-mass
energy of the proton-proton collision, the variation of parton distribution
functions, and the change of factorization scale is investigated.
The sensitivity of this process to the anomalous couplings is examined.
\vspace{1cm}\\
PACS number(s): 14.65.Ha, 14.80.Bn, 12.60.-i
\vspace{3.3cm}
\end{titlepage}

\newpage

\section{Introduction}

The Higgs boson discovery has been one of the most challenging goals
of present high-energy experiments, in particular at the LHC experiments.
From another side, the study of top quark interactions provide a window to new physics
beyond the Standard Model (SM).
Because of the large mass of the top quark, it is maximally coupled to the Higgs boson.
Therefore, its interactions seem to be more sensitive to the mechanism of electroweak symmetry breaking than other
Standard Model particles \cite{werner},\cite{wagner},\cite{beneke},\cite{heinson},\cite{higgsrev},\cite{djouadi}.
Recently, LHC experiments have observed an excess above the expected background, at a mass near 125 GeV,
which is the signal for production of a new particle.
The measured significance for a SM Higgs boson with this mass is 5.8 standard deviations \cite{cms}.

In the top quark sector, one should note that
the top quark couplings still need to be measured
more precisely because it can provide a place to observe any deviations from the SM
predictions. In this work, by adopting an effective Lagrangian approach, the new physics effects are parameterized
in terms of higher dimensional operators constructed from the Standard Model
fields \cite{efflag},\cite{zhang2}:
\begin{eqnarray}
\mathcal{L}_{eff} = \mathcal{L}_{SM}+\sum_{i}\frac{c_{i}O_{i}}{\Lambda^{2}}+...
\end{eqnarray}
where in the above relation $c_{i}$'s are dimensionless coefficients pointing to the strength of the anomalous couplings,
$O_{i}$ are dimension 6 gauge invariant operators containing the Standard Model fields and $\Lambda$
is the new physics scale.
It is notable that due to the excellent agreement between the SM expectations and
the present experimental data, any deviations from the SM are expected
to be small.
Accordingly, the new effective terms must be very small and the interference term
between SM and new effective terms could be big enough to be measured.
With on-shell top, bottom quark and $W$ boson, the most general form of
the $Wtb$ vertex originating from the dimension six operators can be written as follows \cite{kane},\cite{whisnant},\cite{misiak2}:
\begin{eqnarray}\label{eff}
\mathcal{L}_{eff} = -\frac{g}{\sqrt{2}}\bar{b}[\gamma_{\mu}(V_{L}P_{L}+V_{R}P_{R})+\frac{i\sigma_{\mu\nu}q^{\nu}}{m_{W}}(g_{L}P_{L}+g_{R}P_{R})]tW^{\mu}+h.c.,
\end{eqnarray}
where $q$ is the momentum of the involved $W$ boson, $P_{R,L} = \frac{1\pm \gamma_{5}}{2}$, $g$ is the weak coupling.
Within the Standard Model, only the first term with $V_{L} = V_{tb} \simeq 1$ is present and the remaining
terms are anomalous terms which vanish at tree level and can be generated by radiative corrections.
The assumption of CP conservation leads that $V_{L,R},g_{L,R}$ to be real.

There are indirect and direct constraints on the anomalous couplings. The indirect constraints
have been extracted from the measured branching
ratio of $b \rightarrow s\gamma$. In particular, the bounds on $V_{R}$ and $g_{L}$ are tight.
These bounds are stronger than what can be possibly obtained from
studying the top quark production and decay at the LHC. This is because of the fact that the
decay amplitude of B-meson is enhanced by a factor $m_{t}/m_{b}$.
The bounds are presented below \cite{misiak}:
\begin{eqnarray}
-0.0007 < V_{R} < 0.0025~,~ -0.0013 < g_{L} < 0.0004~,~ -0.15 < g_{R} < 0.57
\end{eqnarray}

The recent direct constraints on the anomalous couplings have been obtained using $p\bar{p}$ collision data
corresponding to an integrated luminosity of 5.4 $fb^{-1}$ collected by the D0 detector.
The analysis is based on the $t$-channel and $s$-channel single top quark production cross sections.
The result represents the most stringent direct limits on
anomalous $Wtb$ couplings  which can be found in the following (assuming $V_{tb} \simeq 1$) \cite{d0}:
\begin{eqnarray}
 |V_{R}| < 0.96~,~ |g_{L}| < 0.36~,~ |g_{R}| < 0.24
\end{eqnarray}

As it can be seen, the indirect constraints are much stronger than the
direct ones except for the upper constraint on $g_{R}$. In spite of the direct constraints, the indirect bounds are not symmetric.
In this work, because of the strong limit on $V_{R}$ and for simplicity we do not consider it.
There are several studies on the anomalous $Wtb$ couplings which for example some of them can be found in \cite{rindani},
\cite{boos},\cite{werner2},\cite{larios},\cite{chen}, \cite{moj}, \cite{wtb1},\cite{wtb2},\cite{wtb3},\cite{wtb4},\cite{wtb5}.

\begin{figure}
\centering
  \includegraphics[width=7cm,height=5cm]{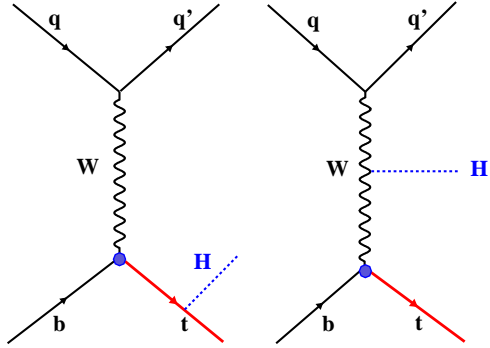}
  \caption{Feynman diagrams contributing to the $t$-channel single top
  plus Higgs production at hadron colliders.}\label{feyn}
\end{figure}

At the LHC, the production of Higgs and single top quark are through three channels:
$t$-channel ($qb \rightarrow q'tH$), $s$-channel ($q\bar{q}' \rightarrow \bar{b}tH$), and
$W$-associated ($gb \rightarrow tW^{-}H$) \cite{maltoni}. Since $t$-channel has the largest cross section
among these channels, we only focus on the $t$-channel.
The possibility of detection a signal for the $t$-channel process (which has the largest cross section) at the LHC, via the Higgs decay into $b\bar{b}$
has been discussed in \cite{maltoni}.
Figure \ref{feyn} shows the Feynman diagrams for the $t$-channel process.

In single top production in the $t$-channel mode ($qb\rightarrow
q't$), the top quarks are produced highly polarized along the
direction of the light-quark in the final state \cite{mahlon},\cite{mahlon2}. The top
quark spin is maximally correlated with the light-quark momentum in
the final state and this correlation can be observed in the angular
distribution of the top decay products.

In this paper, first we study the spin correlation effect for the
production of a Higgs boson plus a single top in the $t$-channel at
the LHC ($qb \rightarrow q'tH$) in the Standard Model. We investigate
the robustness of the spin correlation against variation of the
parton distribution functions (PDFs), variation of the
factorization scale, and change of the center-of-mass energy of
proton-proton collision.
Then we investigate the effect of anomalous $Wtb$ couplings
 as introduced in Eq.\ref{eff}
 on the spin correlation and the production cross section of the
 $qb \rightarrow q'tH$ process.
We also study the change of cross section with the anomalous couplings and its sensitivity to the
 Higgs boson mass.
 The angular correlation distribution of the charged lepton is used to examine
the sensitivity of this process to the anomalous couplings $g_{L}$ and $g_{R}$. All results are presented
at the LHC with the center-of-mass energy of 8 TeV.

The paper is organized as follows. In section 2, the spin correlation in the process of $qb \rightarrow q'tH$
is investigated. In section 3,  we investigate the dependency of the cross section
of the $t$-channel process to the anomalous couplings for different Higgs boson masses as well as
the spin correlation distribution.
We also present the dependence of the rapidity distribution of the light quark to the anomalous couplings. In section 3,
using the charged lepton angular distribution, the sensitivity of the process to the anomalous couplings is examined.
Section 4 is dedicated to conclusions.

\section{Spin Correlation in $qb \rightarrow q'tH$}

It is well-known that for the decay of a top quark into a charged lepton,
a b-quark and a neutrino the angular distributions in the top quark rest frame is given by:
\begin{eqnarray}
\frac{1}{\Gamma}\frac{d\Gamma}{d\cos\theta_{i}} = \frac{1}{2}(1+\alpha_{i}\cos\theta_{i})
\end{eqnarray}
where $\theta_{i}$ is the angle between the top spin and the three-momentum of the
$i$th product. The constants $\alpha_{i}$ are called spin correlation coefficients
or spin analyzing power of the $i$th product which varies from -1 to 1. In the SM at tree level,
$\alpha_{l^{+}} = 1, \alpha_{\nu_{l}} = -0.319$ and $\alpha_{b} = -0.406$.

In the $t$-channel single top production ($qb\rightarrow q't$), top quarks are produced
highly polarized along the light-quark direction in the final state. The corresponding
angular distribution of the charged lepton in the top quark decay is given by:
\begin{eqnarray}\label{angular}
\frac{1}{\Gamma}\frac{d\Gamma}{d\cos\theta_{l}} = \frac{1}{2}(1+\mathcal{P}\alpha_{l}\cos\theta_{l})
\end{eqnarray}
where the angle $\theta_{l}$ is the angle between the charged lepton momentum and the top
spin quantization axis in the top quark rest frame. $\mathcal{P}$ is the spin asymmetry and it is dependent on
the basis. In a basis where top quarks are $100\%$ polarized, $\mathcal{P} = 1$. Choosing the momentum of
the light-quark ($q'$) in the final state (``spectator basis'') as the spin quantization axis leads to high degree
polarization for the top quark ($\mathcal{P} = 97\%$). There is another basis which is called ``beam line'' basis
where the polarization degree is less than the spectator basis \cite{mahlon}, \cite{sullivan},\cite{lev},\cite{Motylinski:2009kt}.

\begin{figure}
\centering
  \includegraphics[width=9cm,height=7cm]{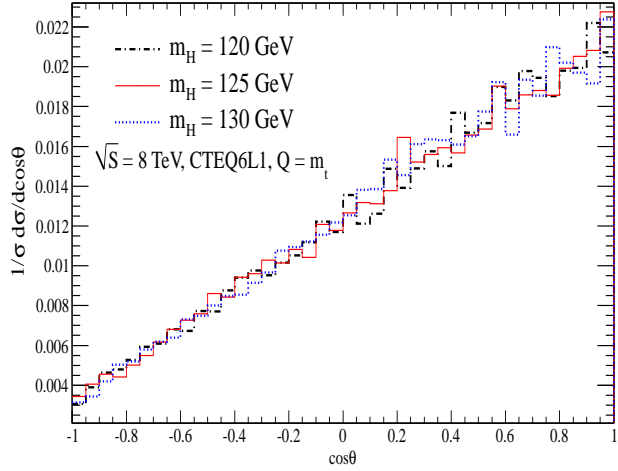}
  \caption{Angular correlation in single top and anti-top for various Higgs boson mass in the spectator basis.
}\label{spinMH}
\end{figure}

Now, we investigate the spin correlation in the process of $qb \rightarrow q'tH$ and
try to obtain the the degree of polarization in the spectator basis for this process.
This has been calculated using tree-level matrix elements generated
by CompHEP \cite{comphep}
convoluted with the parton distribution
function set CTEQ6L \cite{cteq}. We used $m_{t} = 173.1$ GeV and $\Gamma_{t} = 1.4$ GeV.
We consider the top quark decays into a $W$-boson and a $b$-quark and then $W$-boson
decays into a charged lepton and a neutrino. Figure \ref{spinMH} depicts the angular correlation
distribution of the charged lepton in the spectator basis for three values of the Higgs boson mass.

As it can be seen in Figure \ref{spinMH}, similar to the process $qb\rightarrow tq'$, a linear
behavior is observed. Thus, we fit a linear function of $az+b$, where $z = \cos\theta$, to the angular distribution
to extract the parameters $a$ and $b$. Comparing with Eq. \ref{angular}, $a = \mathcal{P}\alpha_{l}$. If we assume
$\alpha_{l} = 1$ then $a = \mathcal{P}$. For the Higgs boson mass $m_{H} = 125$ GeV, we obtained
$\mathcal{P}$ to be $76.4\%$. Variation of the Higgs boson mass within 5 GeV leads to a negligible
change of $3\%$ in the spin asymmetry.
It is interesting to note that with respect to the single top $t$-channel process $qb\rightarrow q't$, in the
spectator basis, the spin asymmetry in this process is degraded around $22\%$.

\begin{figure}
\centering
  \includegraphics[width=7cm,height=5cm]{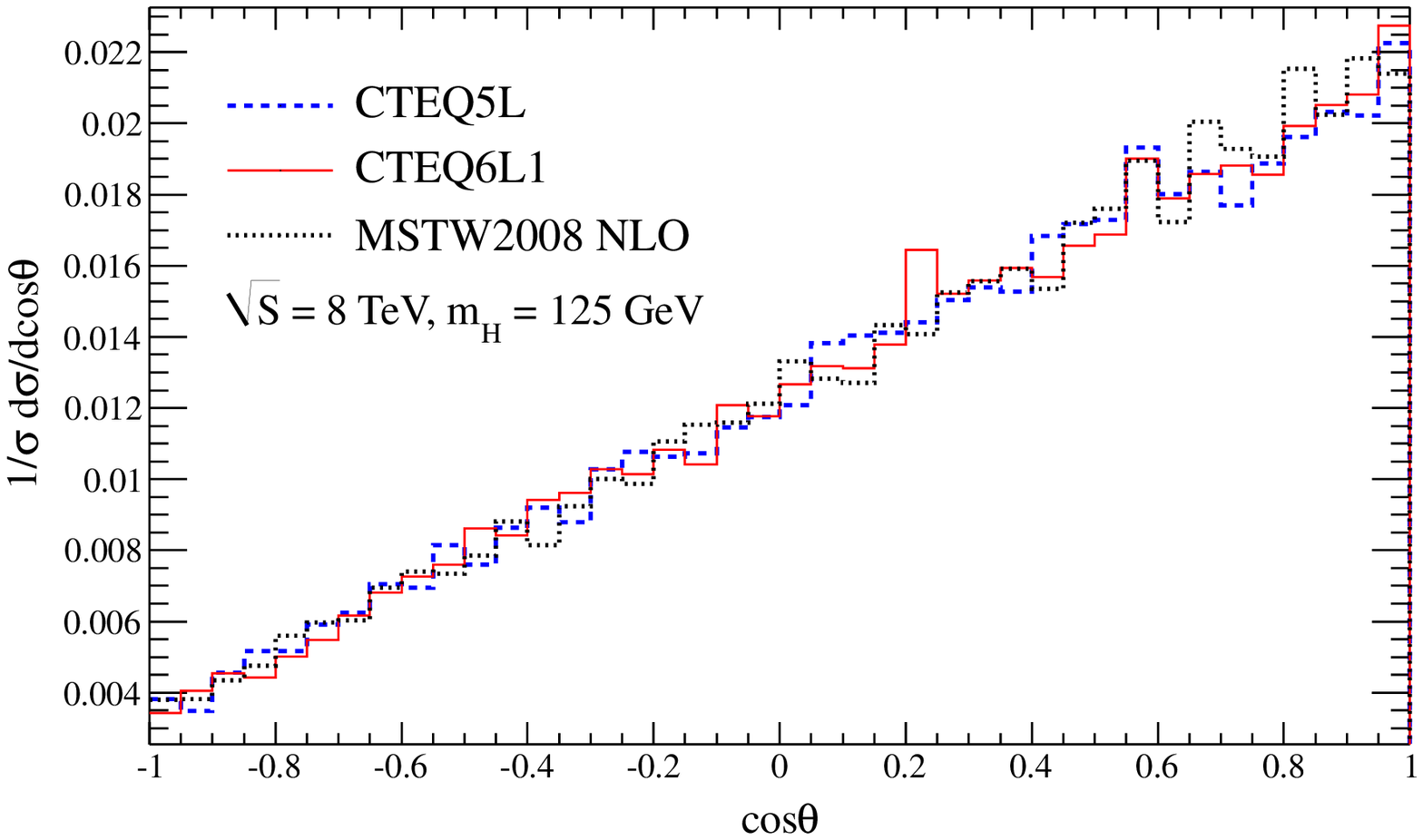}
  \includegraphics[width=7cm,height=5cm]{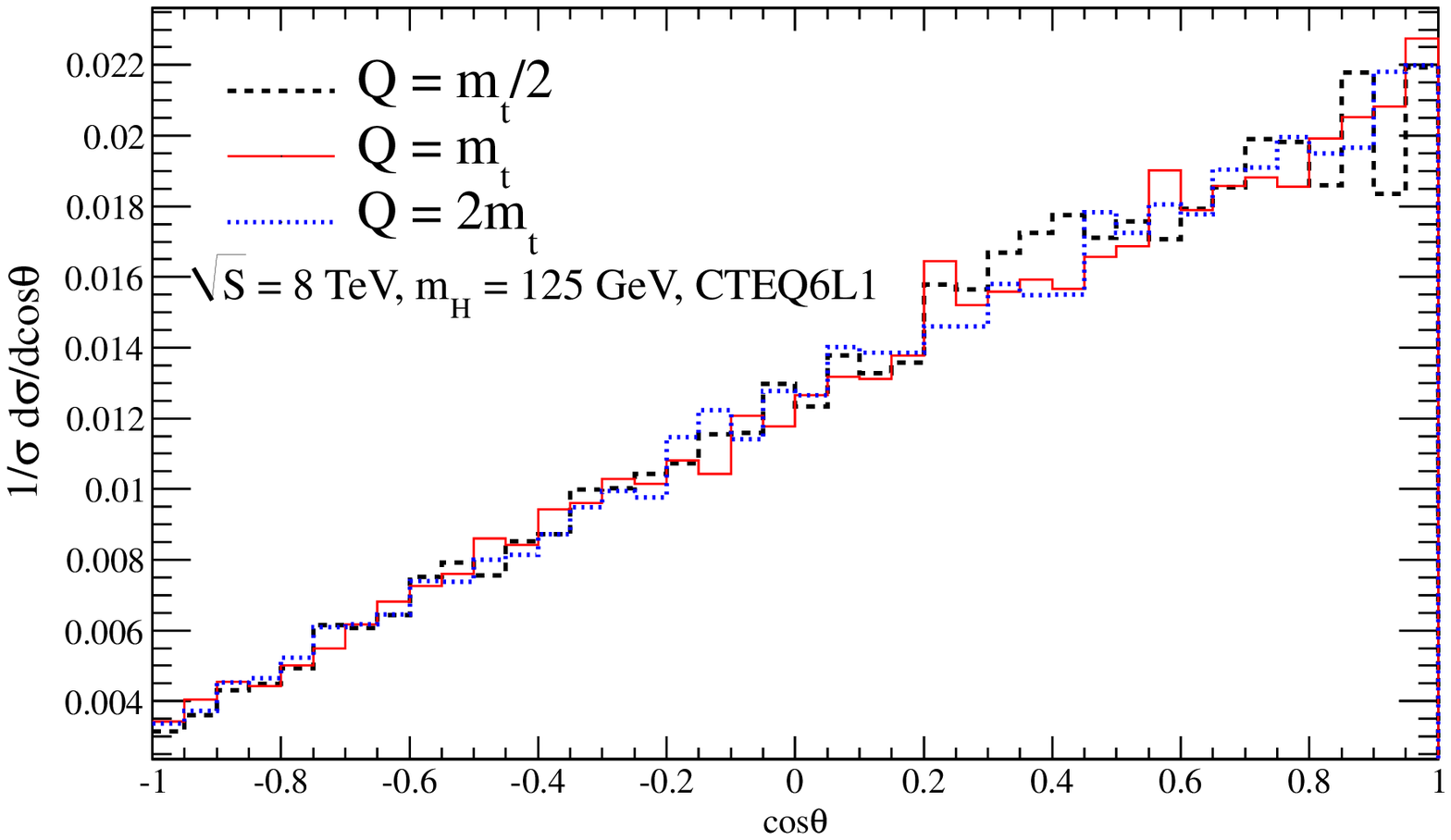}
  \caption{Angular distribution using three PDF sets (left) and for three values of factorization scale with
a given PDF set (right).
}\label{pdf}
\end{figure}

Now, we check the sensitivity of the polarization degree to the change of
proton parton distribution functions and the variation of the factorization scale.
In order to examine the influence of the choice of PDF on the polarization degree $\mathcal{P}$,
we compare the angular distributions obtained with CTEQ6L1, CTEQ5L, and MSTW2008nlo. Figure \ref{pdf} (left plot)
shows the results. As it can be seen, no considerable differences is observed when we use
different set of PDFs. In the spectator basis, the maximum differences in $\mathcal{P}$ between using three PDF sets
is $1.5\%$.
Figure \ref{pdf} (right plot) shows the angular distribution for a given PDF set (CTEQ6L1) with
various factorization scales $\mu_{F} = m_{t}/2,m_{t}, 2m_{t}$. According to figure, the polarization
does not receive any substantial changes when different factorization scales are used in the
calculations. The difference is at the level of $1\%$.

\begin{figure}
\centering
  \includegraphics[width=9cm,height=7cm]{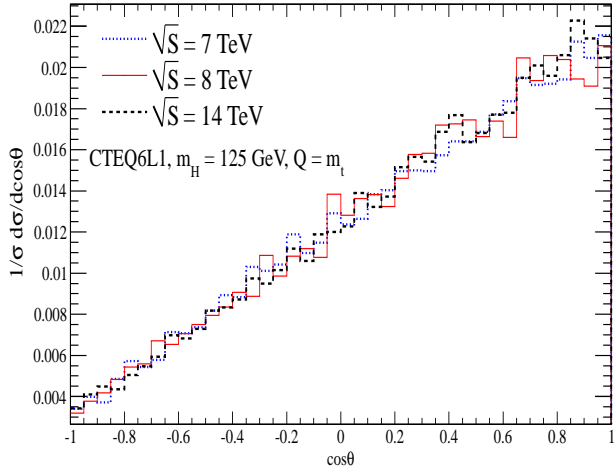}
  \caption{Angular distribution for various center-of-mass energies in the spectator basis at the LHC.
}\label{cme}
\end{figure}

The Large Hadron Collider has started its operation with the center-of-mass energy of 7 TeV and
around 5 fb$^{-1}$ of data collected by the experiments. Large amount of data (around 20 fb$^{-1}$)
has been collected at 8 TeV center-of-mass energy in 2012. The plan is to increase the energy to
larger value in the upgrade of the machine. Thus, it is useful to check the dependency of the
polarization to different center-of-mass energies at the LHC. Figure \ref{cme} shows the
angular distribution for three center-of-mass energies of the proton-proton collisions in the
spectator basis. The degree of polarization is stable against variation of the center-of-mass
energies. Note that the center-of-mass energies are chosen large enough above the
threshold of production.

\section{Associated Production of a Single Top and a Higgs Boson with Anomalous $Wtb$ interactions}

In this section we focus on the effect of the anomalous $Wtb$
interactions on the $t$-channel ($qb \rightarrow q'tH$) single top
quark production in association with a Higgs boson.
This has been calculated using tree-level matrix elements generated
by CompHEP \cite{comphep}
convoluted with the parton distribution
function set CTEQ6L1 \cite{cteq}. The top quark mass has been taken $m_{t} = 173.1$ GeV and its width is
1.4 GeV.

With the general effective Lagrangian for the $Wtb$ coupling, the cross section of the
$t$-channel can be parameterized as the following:
\begin{eqnarray}
\sigma = \sigma_{SM}(V_{tb}^{2}+ A_{1} \times g_{L}^{2} + A_{2} \times g_{R}^{2}
+ A_{3} \times V_{tb} g_{L}+ A_{4} \times V_{tb} g_{R} + A_{5} \times g_{L}g_{R})
\end{eqnarray}
where we have taken $V_{L} = V_{tb} \simeq 1$. $A_{i}$ coefficients determine the dependence of the
cross section to the anomalous couplings. These coefficients are dependent on the proton parton
distribution functions (PDF's) as well as the $Q$-scale and other parameters such as masses of the top
and bottom quarks. It should be noted that $A_{i}$ factors are different for top quark and anti-top
quark production cross sections.

In Table \ref{coefficients}, the $A_{i}$ coefficients are listed for three different masses of the
Higgs boson. The cross section of the single top (anti-top) process within the SM ($\sigma_{SM}$)
is 10.627 (4.634) fb for $m_{H} = 125$ GeV/c$^{2}$.

\begin{table}\caption{
The coefficients of anomalous terms appear in the cross section of the $t$-channel process
for three values of the Higgs boson mass for single top and single anti-top quark separately.}
\begin{center}
\begin{tabular}{c||c|c|c|c||c|c|c|c}
  \hline
  $m_{H}$ [GeV/c$^{2}$] & $A_{1}^{t}$& $A_{2}^{t}$&$A_{3}^{t}$&$A_{4}^{t}$& $A_{1}^{\bar{t}}$& $A_{2}^{\bar{t}}$&$A_{3}^{\bar{t}}$&$A_{4}^{\bar{t}}$  \\ \hline
           124     &14.178 & 7.603 & -2.645  &-0.0626 & 8.669 & 15.553& -1.250&-0.0264 \\
           125     & 14.041& 7.500 & -2.623  &-0.0644 & 8.634 &15.507& -1.237 &-0.0277 \\
           127     & 13.963& 7.422 & -2.605  &-0.0464 & 8.488 & 15.312&-1.221 &-0.0271 \\

  \hline
\end{tabular}\label{coefficients}
\end{center}
\end{table}

As it can be seen in Table \ref{coefficients}, the coefficients of $g_{L}^{2}$ and
$g_{R}^{2}$ terms are greater than the other terms. This has been expected from $q_{\nu}$
factor in the interaction which has an enhancement effect on the cross section.
Similar situation occurs in the SM single top production
with anomalous couplings \cite{aguilar2}. According to Table \ref{coefficients},
the coefficient of the interference term $V_{tb}g_{L}$ is of order of unity
and could not be neglected. While the coefficients of terms $V_{tb}g_{R}$
and $g_{L}g_{R}$ are very small, at the order of $10^{-2}$ and $10^{-3}$, respectively.
Therefore, it might not be a precise assumption to take only one non-zero anomalous coupling at a time.

In comparison between the coefficients of single top and single anti-top quarks
in Table \ref{coefficients}, it is notable that the coefficient of $g_{L}^{2}$ term
in single top is larger than single anti-top while the situation is vice versa
for the $g_{R}^{2}$ term. Therefore, the ratio of the top to anti-top cross section is more
sensitive to $g_{L}$.

From Table \ref{coefficients}, it can be seen that the coefficients are
sensitive to the Higgs boson mass. However, the sensitivity is not significant.
The coefficients change by less than $2\%$ when the Higgs boson mass
varies within 3 GeV.

\begin{figure}
\centering
  \includegraphics[width=7cm,height=5cm]{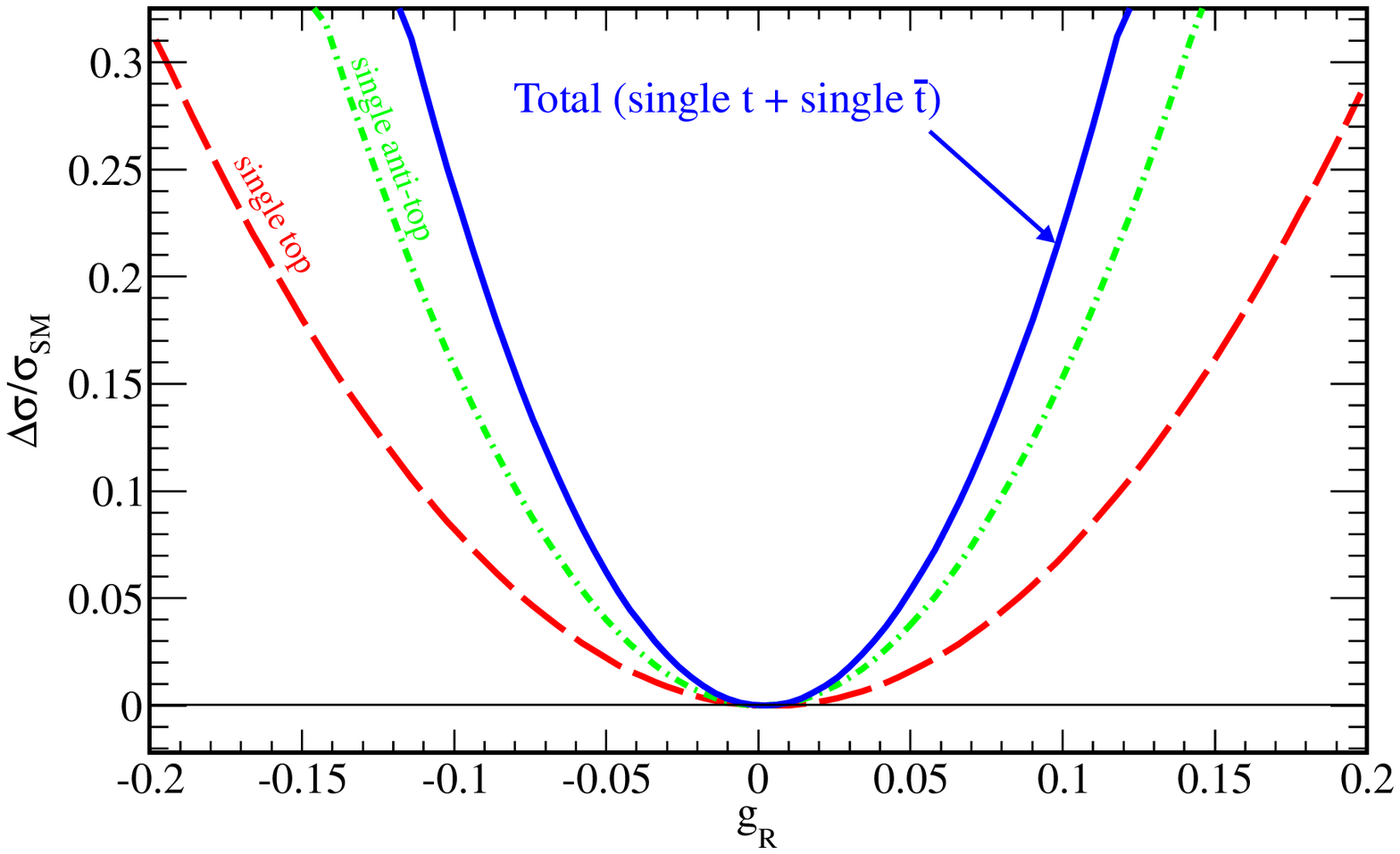}
  \includegraphics[width=7cm,height=5cm]{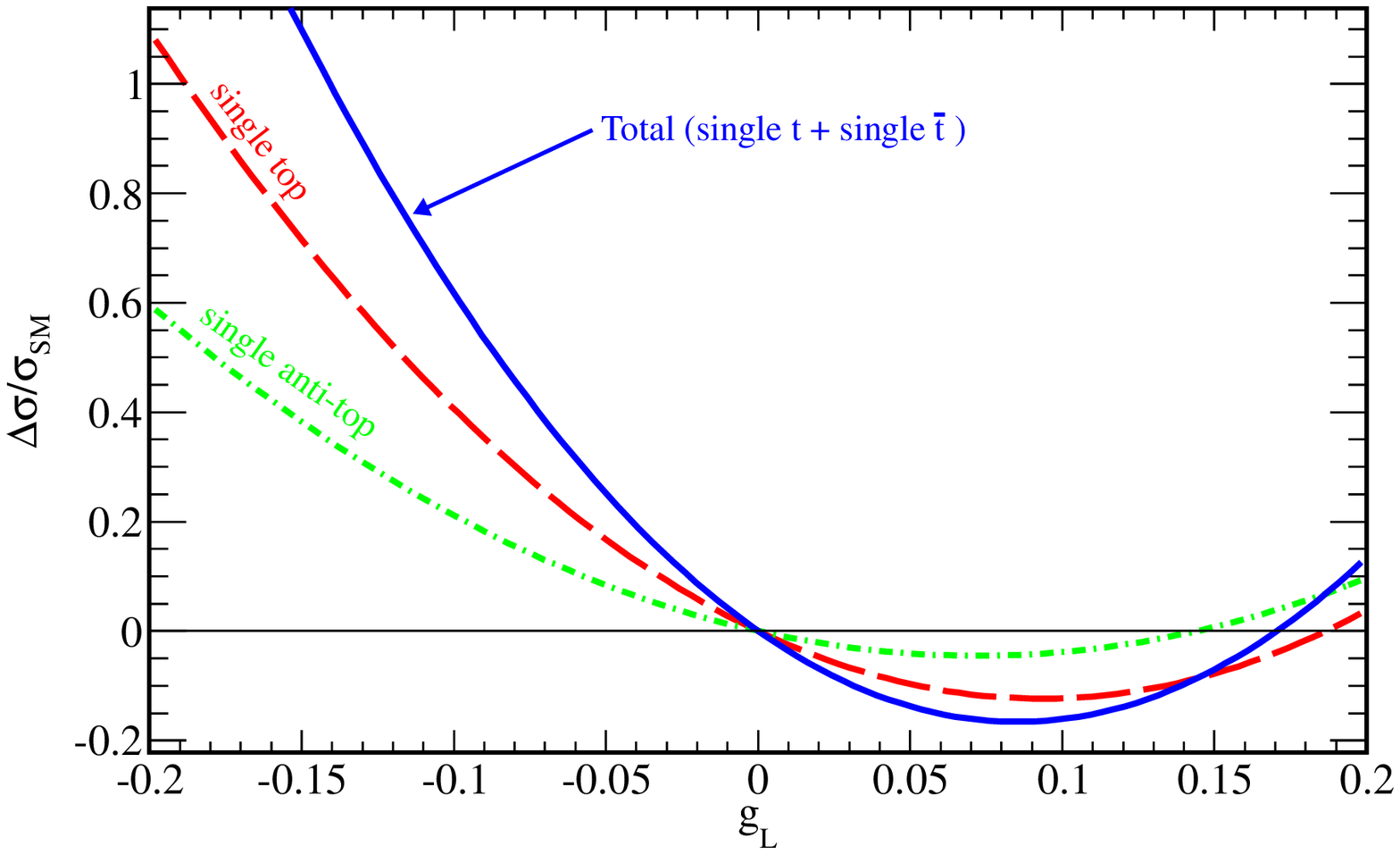}
  \caption{The relative correction coming from the anomalous
    interactions to the cross section for production
    of Higgs plus single top in the $t$-channel at the LHC ($pp$,
    $\sqrt{s} = 8$ TeV). The dependence of the change of cross section
    on $g_{R}$ (left) and on $g_{L}$ (right) for $m_{H} = 125$ GeV.
}\label{gr}
\end{figure}

Fig. \ref{gr} depicts the relative change of the cross section for production
of a Higgs boson plus a single top in the $t$-channel at the LHC with the
center-of-mass energy of $\sqrt{s} = 8$ TeV in the presence of the anomalous $Wtb$
interactions. The dependence of the change of cross section
on $g_{R}$ (left) and on $g_{L}$ (right) is for $m_{H}=125$ GeV.

By comparing both plots in Fig. \ref{gr}, one can observe that the
cross section is more sensitive to $g_{L}$ (left-handed with tensor structure)
than $g_{R}$ (right-handed with tensor structure). For example, the
contribution that the cross section can receive from $g_{L} = 0.05$ reaches
up to $20\%$ while from $g_{R} = 0.05$ is $6\%$.
According to Fig. \ref{gr} the anomalous term of $g_{L}$ could be destructive and reduces the cross section from
its Standard Model value up to around $20\%$ while $g_{R}$ always increases the cross section.

One of the interesting aspect of the $t$-channel process is that the light-quark
in the final state tend to fly in the forward/backward region of the detector.
The presence of a light-jet in the forward or backward region is related to the cross section behavior
as a function of the momentum of the virtual $W$-boson exchanged in the $t$-channel.
Similar to the $t$-channel single top production (without Higgs boson in the final state),
the region $-q^{2} \leq m_{W}^{2}$ dominates \cite{singletop}.

\begin{figure}
\centering
  \includegraphics[width=10cm,height=7cm]{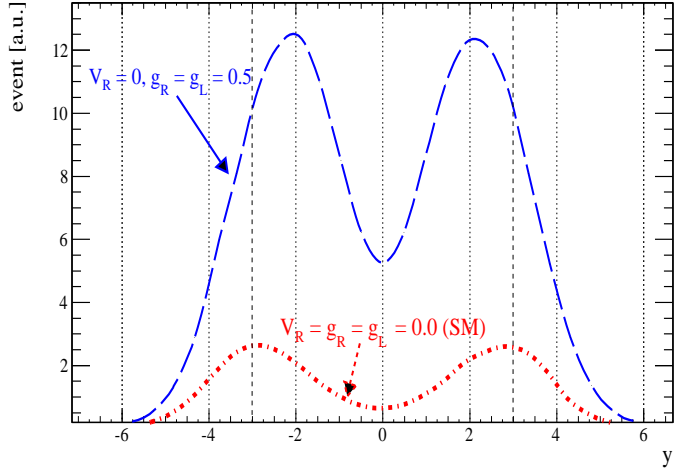}
  \caption{Rapidity distribution for the light quark in the final-state in the t-channel at the LHC.
}\label{eta}
\end{figure}

In Fig. \ref{eta} we present the rapidity distribution of the light-quark
in the final-state of the $t$-channel in the SM framework and in
the presence of anomalous couplings with for example $g_{L} = g_{R} =
0.5$. As it can be seen in Fig. \ref{eta}, the light quark
tends to reside at large rapidities, peaking at around the value
of 3 in the SM framework which is corresponding to an angle of around $5.7^{\circ}$. The interesting point is that
the presence of anomalous couplings leads the light quark to be more
central. For example, when $g_{L} = g_{R} =0.5$ the light quark
rapidity distribution is peaking at around 2 (corresponding to an angle of $15.4^{\circ}$), one unit below the SM case.
Therefore, in the presence of the anomalous couplings the spectator quark tends
to fly more central and it affects the spin correlation and angular distribution which
have been discussed in the previous section. In the next section, we use the angular distribution to
examine the sensitivity of the process to the anomalous couplings.

\section{Sensitivity to Anomalous Couplings}

As we showed in section 2, there is a correlation between the spin of the top
quark and the three-momentum of the light-quark in the top quark rest frame. The
polarization degree in $qb \rightarrow q'tH$ process in the spectator basis is around $76\%$
for $m_{H} = 125$ GeV. The presence of anomalous couplings in the vertex of $Wtb$
changes this polarization degree. In Table \ref{xxx}, the degree of
polarization is shown for various values of $g_{L}$ and $g_{R}$ in the
spectator basis. There are interesting observations:
\begin{itemize}
\item{When we change $g_{R}$ in the range of 0.0 to 0.1, an
  increase of around $4\%$ is seen with respect to the SM
  expectation and when we move in the negative direction from 0.0 to
  -0.1, we face with a decrease of $3-4\%$. Therefore, the polarization
  degree is weakly sensitive to $g_{R}$.}
\item{For the case that $g_{L}$ varies from 0.0 to 0.1 , there
  is an increase of $3-4\%$ in the polarization degree while variation
  from of $g_{L}$ from 0.0 to -0.1 leads to a significant deviation of
  $27\%$. The polarization degree is strongly sensitive to $g_{L}$
  when $g_{L} < 0.0$.}
\end{itemize}

In summary, the angular correlation of the charged lepton is more
sensitive to $g_{L}$ than $g_{R}$ and in particular to negative values
of $g_{L}$. We use the charged lepton angular distribution to examine
the sensitivity of this process to the anomalous couplings.

\begin{table}\caption{
The coefficients of anomalous terms appear in the cross section of the $t$-channel process
for three values of the Higgs boson mass for single top and single anti-top quark separately.}
\begin{center}
\begin{tabular}{|c|c|c|}
  \hline
  $g_{L}$ & $g_{R}$ & Degree of Polarization   \\ \hline
     0.1   &  0.0   &  0.7938  \\
     0.0   &  0.1   &  0.7963  \\
     0.1   &  0.1   &  0.8091  \\
    -0.1   & -0.1   &  0.5392  \\
     0.2   & 0.2    &  0.6030  \\
    -0.2   & -0.2   &  0.3416  \\
    -0.1   &  0.0   &  0.5536  \\
     0.0   &  -0.1  &  0.7497  \\
  \hline
\end{tabular}\label{xxx}
\end{center}
\end{table}

To obtain a realistic estimation of the sensitivity of the single top in association
with the Higgs boson in $t$-channel, one has to consider all backgrounds, selection cuts,
and detector effects. However, a comprehensive analysis taking into account all backgrounds
and detector effects is beyond the scope of this paper and must be performed by the experimental
collaborations. \\
In this study, we use $\chi^{2}$ criterion from the charged lepton angular distribution
to estimate the sensitivity to the anomalous couplings:
\begin{eqnarray}
\chi^{2}(g_{L},g_{R}) = \sum_{i=bins} (\frac{N_{i}^{new}(g_{L},g_{R})-N_{i}^{SM}}{\Delta_{i}})^{2}
\end{eqnarray}
where $N_{i}^{new}$ is the number of expected events in $i$th bin of
the charged lepton distribution in the presence of anomalous couplings.
The same as \cite{maltoni}, we consider the leptonic decay of the $W$-boson form the top quark
and the decay of the Higgs boson to $b\bar{b}$. Since for a Higgs boson lighter than 130 GeV/c$^{2}$,
the dominant decay mode is $b\bar{b}$, we will have more statistics
with this choice. Therefore, our final state consists of
$l^{\pm}+$Missing Transverse Energy(MET)+$3b$jet$+$ one light forward jet.
As discussed in \cite{maltoni} and in a more recent paper
\cite{Farina}, the main backgrounds are $t\bar{t}$, $t\bar{t}j$, $tZj$
with $Z\rightarrow b\bar{b}$, and $tb\bar{b}j$. Following the set of
cuts proposed in \cite{Farina} leads to significant suppression of the
backgrounds. In \cite{Farina}, the acceptance cuts has been taken
similar to the ATLAS analysis of $t\bar{t}H$ \cite{atlastt}.
The transverse momenta of the charged lepton and the bjet are required
to be greater than 25 GeV with $|\eta_{l,b}|<2.5$. The light jet must
have $p_{T} > 30$ and $|\eta| < 5.0$. It is also required that $\Delta
R_{ij} > 0.4$. Where $\Delta R_{ij} =
\sqrt{(\eta_{i}-\eta_{j})^{2}+(\phi_{i}-\phi_{j})^{2}}$, this
requirement is applied on all objects. No cut on MET is applied.
In the $\chi^{2}$ definition, $\Delta_{i}$ denotes the uncertainties in each bin of the
angular distribution and is defined as: $\Delta_{i} = N_{i}^{SM}\sqrt{\delta_{stat}^{2}+\delta_{sys}^{2}}$.
After imposing the kinematic cuts, the charged lepton angular
distributions are distorted. A significant drop in the region of
$\cos\theta \sim 1$ appears which is mainly originating from the
$p_{T}$ and $\Delta R$ cuts. This behavior has been already observed
in the SM single top production $qb\rightarrow tq'$.
It is important to note that the main characteristic feature of signal
in the charged lepton angular distribution, which is
a distinct slope, remains. The background has a flat distribution
even after the cuts are imposed. In the SM single top production
$qb\rightarrow tq'$, this feature remains after full detector
simulation and has been observed in real data too \cite{cmsst}.

To get a rough estimate of the limits on the anomalous couplings, we consider an integrated luminosity of 25 fb$^{-1}$ of data.
The $68\%$ C.L. interval for $g_{L}$ and $g_{R}$ are $[-0.32,0.65]$ and $[-0.74,0.74]$, respectively.
Indeed, these are rough estimations and in the realistic case detector effects, all backgrounds must be taken into account.
Because of smaller statistics, the limits that obtained from single top production in association with the Higgs boson process are looser than the
ones that can be obtained from for example $t$-channel single top \cite{aguilar2}. However, the same as \cite{aguilar2} where
all single top channels have been combined, the combination of this channel with
other sensitive channels can improve the constraints on the anomalous terms
\cite{aguilar1},\cite{aguilar2},\cite{mojtaba}.

\begin{figure}
\centering
  \includegraphics[width=8cm,height=6cm]{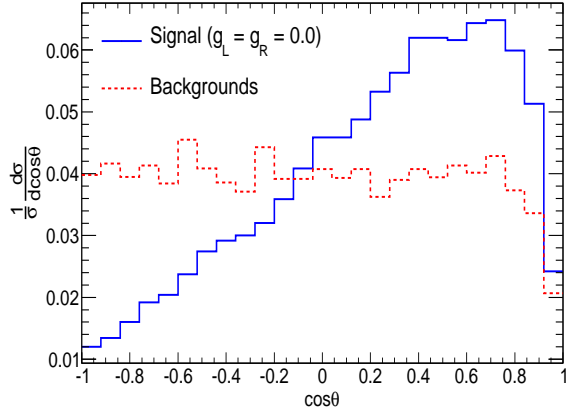}
  \caption{The $\cos\theta$ distributions of the charged lepton in top
  decay for signal and backgrounds after the cuts.}\label{cut}
\end{figure}

\section{Conclusions}

In this paper, we studied the single top quark production
in association with a Higgs boson at the LHC, namely $qb\rightarrow
tq'H$.
In particular, we concentrated on the spin correlation effects in this
process within the SM framework and in presence of the anomalous $Wtb$
couplings. We have found that in the SM, in the process of $qb
\rightarrow tq'H$, the top quarks are $\sim 76\%$ polarized in the
direction of the light-quark $q'$ in the final state.
It was shown the polarization degree deviates negligibly by changing
the PDF set and the center-of-mass energy of the collision (7,8,14 TeV).
The dependence of the spin correlation on the factorization scale has
been found to be $1\%$.

We found that the amount of change in cross section with the anomalous couplings is around $2\%$ when
the Higgs boson mass varies 3 GeV/c$^{2}$. Variation of the Higgs
boson mass has a weak effect on the top polarization degree.
The cross section was found to be
strongly sensitive to the anomalous couplings $g_{L},g_{R}$. The
$g_{R}$ anomalous coupling contribution to the cross section is always
constructive. On the contrary, the $g_{L}$ anomalous coupling can
contribute to the cross section destructively ($0.0 < g_{L} < 0.2$).
It is interesting to note that the cross section is more sensitive to
$g_{L}$ than $g_{R}$.

Since the charged lepton in the top quark decay is maximally
correlated with the top quark spin, we performed a parton level study
to examine the anomalous $Wtb$ couplings in this process by focusing
on the top quark leptonic decay and the Higgs boson decays into $b\bar{b}$.
After imposing a set of kinematic cuts, we showed that the angular
distribution of the charged lepton remains as a sloped line (except
for the region of $\cos\theta \sim 1$, which is due to the $p_{T}$ and
$\Delta R$ cuts). The flatness of background shape remains after the
cuts. We examined the sensitivity to the anomalous couplings using
the charged lepton angular distribution. With 25 fb$^{-1}$ of data,
the $68\%$ C.L. interval for $g_{L}$ and $g_{R}$ have been found to be $[-0.32,0.65]$ and $[-0.74,0.74]$.

\vspace{0.5cm}
{\bf Note added:} When this work was being completed, a partly similar work appeared in \cite{Agrawal} where the
dependence of cross section on the anomalous couplings has been studied.

\vspace{0.1cm}
{\bf Acknowledgments}\\
S.~F.~Taghavi is grateful to Bonyad Melli Nokhbegan.
M.Mohammadi Najafabadi is grateful to CompHEP authors
in particular E. Boos, L. Dudko, and A. Sherstnev. The authors are
also thankful to S. Khatibi for her help in extracting the limits on 
anomalous couplings.

\end{document}